\documentstyle[aps,preprint,prb]{revtex}
\begin{document}
\tightenlines
\def\vk{\vec k} 
\def\br{{\bf r}}
\title{\bf On the Mooij Rule }
\author{Mi-Ae Park }
\address{Department of Physics,  University of Puerto Rico at Humacao,\\
 Humacao, PR 00791}
\author{Kerim Savran and Yong-Jihn Kim }
\address{Department of Physics,  Bilkent University,\\
 06533 Bilkent, Ankara, Turkey}
\maketitle
\begin{abstract}

Weak localization leads to the same correction to both the conductivity
and the electron-phonon coupling constant $\lambda$ (and $\lambda_{tr}$). 
Consequently the temperature dependence of the (thermal) electrical
resistivity is decreasing as the conductivity is decreasing due to
weak localization, which results in the decrease of the temperature coefficient
of resistivity (TCR) with increasing the residual resistivity.  
When $\lambda$ is approaching zero, only residual resistivity part remains  
and gives rise to the negative TCR.
Accordingly, the Mooij rule is a manifestation of weak localization 
correction to the conductivity and the electron-phonon interaction. 
This study may provide a new means of probing the phonon-mechanism in exotic
superconductors.

\end{abstract}

\vspace{1pc}

\noindent

\vfill\eject
\section{\bf Introduction} 

Although weak localization has greatly deepened our understanding of the 
normal state of disordered metals,$^{1,2,3}$ its effect
on superconductivity and electron-phonon interaction has not been
understood well.$^{2}$ Recently, it has been shown that weak localization
leads to the same correction to the conductivity and the phonon-mediated
interaction.$^{4,5}$ It is then anticipated that the electron-phonon interaction
will also be influenced strongly by weak localization. 
For instance, phonon-limited electrical resistance,  attenuation of a sound 
wave, thermal resistance, and a shift in phonon frequencies may change
due to weak localization.$^{6}$

In fact, the Mooij rule$^{7}$ in strongly disordered metallic systems seems to 
be a manifestation of the effect of weak localization on the electron-phonon 
interaction and the conductivity. In early seventies,
Mooij found a correlation between the residual resistivity and the temperature
coefficient of resistivity (TCR). In particular, TCR is decreasing with
increasing the residual resistivity. Then it becomes negative above 
$150\mu\Omega cm$. 
There are already several theoretical works on this problem. 
Jonson and Girvin$^{8}$ performed numerical calculations for an Anderson model on a Cayley tree and found that the adiabatic phonon approximation 
breaks down in the high-resistivity regime producing the negative TCR.
Imry$^{9}$ pointed out the importance of incipient Anderson localization 
(weak localization) in the resistivities of highly disordered metals.
He argued that when the inelastic mean free path, $\ell_{ph}$, is
smaller than the coherence length, $\xi$, the conductivity increases
with temperature like $\ell_{ph}^{-1}$ and thereby leads to the negative
TCR. On the other hand, Kaveh and Mott$^{10}$ generalized the Mooij rule.
Their results are as follows: The temperature dependence of the conductivity 
of a disordered metal as a function of temperature changes slope due to 
weak localization effects, and if interaction effects are included, 
the conductivity changes its slope three times. 
G\"otze, Belitz, and Schirmacher$^{11,12}$ introduced a theory with
phonon-induced tunneling. There is also the extended Ziman theory.$^{13}$ 

In this paper, we propose an explanation of the Mooij rule based
on the effect of weak localization on the electron-phonon interaction.
If we assume the decrease of the electron-phonon interaction
due to weak localization, we can understand the decrease of TCR with 
increasing the residual resistivity.
The negative TCR is therefore due to weak localization correction to
the Boltzmann conductivity, since when TCR is approaching zero there is no 
temperature-dependent resistivity left. (This latter point is similar to  
Kaveh and Mott's interpretation.$^{10}$) Matthiessen's rule seems to 
remain intact to a large extent even in the highly disordered systems.
In Sec. II, we briefly describe the Mooij rule. In Sec. III, weak localization 
correction to the electron-phonon coupling constant $\lambda$ and 
$\lambda_{tr}$ is calculated. A possible explanation of the Mooij rule 
is given in Sec. IV, and its 
implication is briefly discussed in Sec. V. 
In particular, this study may provide a means to probe the phonon-mechanism
in exotic superconductors.

\section{The Mooij Rule}

According to Matthiessen's rule, resistivity $\rho(T)$ caused by static and 
thermal disorder is additive, i.e.,
\begin{equation}
\rho(T)=\rho_{o}+\rho_{ph}(T),
\end{equation}
where $\rho_{ph}$ is mostly due to electron-phonon scattering.
Mooij found (at high temperatures) that the size and sign of the temperature 
coefficient of resistivity (TCR) in many disordered systems correlate with its 
residual resistivity $\rho_{o}$ as follows:

\begin{eqnarray}
d\rho/dT&>&0  \quad \rm{if} \quad \rho_{o}<\rho_{M}\nonumber\\
d\rho/dT&<&0  \quad \rm{if} \quad \rho_{o}>\rho_{M}.
\end{eqnarray}
Thus, TCR changes sign when $\rho_{o}$ reaches the Mooij resistivity $\rho_{M}\cong 150\mu\Omega cm$. 
Figure 1 shows the temperature coefficient of resistance $\alpha$ versus 
resistivity for transition-metal alloys  obtained by Mooij. 
It is clear $\alpha$ (and TCR) is correlated with the residual  resistivity. 
Note that above $150\mu\Omega cm$ most $\alpha$'s are negative.
Figure 2 shows the resistivity as a function of temperature for pure Ti 
and  TiAl alloys containing 3, 6, 11, and 33\% Al. TCR is decreasing
as the residual resistivity is increasing.
For TiAl alloy with 33\% Al shows the negative TCR.
Since this behavior is generally found in strongly disordered metals and 
alloys, amorphous metals, and metallic glasses, it is called the Mooij rule.
However, the physical origin of this rule has remained unexplained until now.

\section{\bf Weak Localization Correction to Electron-Phonon Interaction} 

Since the electron-phonon interaction in metals gives rise to both the (high 
temperature) resistivity and superconductivity, these properties are
closely related, which was noticed by many workers.$^{14-17}$
In this Section, we show that weak localization leads to the
same correction to the conductivity and the electron-phonon
coupling constant $\lambda$ and $\lambda_{tr}$.

\subsection{High Temperature resistivity}

At high temperatures, the phonon limited electrical resistivity is given 
by$^{17}$
\begin{eqnarray}
\rho_{ph}(T)&=&{4\pi mk_{B}T \over ne^{2}\hbar}\int
{\alpha_{tr}^{2}F(\omega)\over \omega}d\omega,\nonumber\\
&=& {2\pi mk_{B}T \over ne^{2}\hbar}\lambda_{tr},
\end{eqnarray}
where $\alpha_{tr}$ includes an average of a geometrical factor
$1-cos\theta_{\vk\vk'}$ and $F(\omega)$ is the phonon density of states.   
On the other hand, in the strong-coupling theory of superconductivity,$^{18,19}$
the electron-phonon coupling constant is defined by$^{19}$
\begin{eqnarray}
\lambda=2\int{\alpha^{2}(\omega)F(\omega)\over \omega}d\omega. 
\end{eqnarray}
Assuming $\alpha_{tr}^{2}\cong\alpha^{2}$, we obtain
\begin{eqnarray}
\rho_{ph}(T)&=&{2\pi mk_{B}T \over ne^{2}\hbar}\lambda_{tr}\nonumber\\
&\cong& {2\pi mk_{B}T \over ne^{2}\hbar}\lambda.
\end{eqnarray} 
Consequently the electron-phonon coupling constant $\lambda$ determines also
the size and sign of TCR.
Table I shows the comparison of $\lambda_{tr}$ and $\lambda$ for various
materials.$^{20,21}$
The overall agreement between $\lambda_{tr}$ and $\lambda$ is impressive. 

\subsection{Weak localization correction to $\lambda$ and $\lambda_{tr}$}

Now we need to calculate the electron-phonon coupling constant $\lambda$
for highly disordered systems.
We follow the approach by Park and Kim.$^{5}$
(For simplicity we consider an Einstein model with frequency $\omega_{D}$). 
Note that  $\lambda$ can be written as$^{19}$
\begin{eqnarray}
\lambda&=&2\int{\alpha^{2}(\omega)F(\omega)\over \omega}d\omega \\
&=&N_{o}{<I^{2}>\over M<\omega^{2}>},
\end{eqnarray}
where $M$ is the ionic mass and $N_{o}$ is the electron density of states 
at the Fermi level.  
$<I^{2}>$ is  the average over the Fermi surface of the 
square of the electronic matrix element and  
$<\omega^{2}>=\omega_{D}^{2}$.
In the presence of impurities, weak localization leads to a correction to 
$\alpha^{2}$ or $<I^{2}>$, (disregarding the changes of $F(\omega)$ and $N_{o}$). 

The equivalent electron-electron potential in the 
electron-phonon problem is given by,$^{22,23}$
\begin{equation}
V(x - x') \rightarrow {I_{o}^{2}\over M\omega_{D}^{2}} D(x-x'),
\end{equation}
where  $x=({\bf r},t)$ and $I_{o}$ is the electronic matrix element for the plane wave states. 
The Fr\"ohlich interaction at finite temperatures is then obtained by
\begin{eqnarray}
V_{nn'}(\omega, \omega')&=& 
{I_{o}^{2}\over M\omega_{D}^{2}} \int\int d{\bf r}d{\bf r'}
\psi_{n'}^{*}({\bf r}) \psi_{\bar{n}'}^{*}({\bf r'})D({\bf r}-{\bf r'},\omega-\omega')
\psi_{\bar{n}}({\bf r'}) \psi_{n}({\bf r})\nonumber\\
&=& 
{I_{o}^{2}\over M\omega_{D}^{2}} \int|\psi_{n'}({\bf r})|^{2} |\psi_{n}({\bf r})|^{2}d{\bf r}{\omega_{D}^{2}\over
\omega_{D}^{2}+(\omega-\omega')^{2}}\nonumber\\
&=& V_{nn'} {\omega_{D}^{2}\over \omega_{D}^{2}+(\omega-\omega')^{2}},
\end{eqnarray}
where$^{23}$ 
\begin{eqnarray}
D({\bf r}-{\bf r'},\omega-\omega')&=&\sum_{\vec q}{\omega_{D}^{2}\over (\omega-\omega')^{2}+\omega_{D}^{2}}
e^{i{\vec q}\cdot({\bf r}-{\bf r'})}\nonumber\\
&=& {\omega_{D}^{2}\over (\omega-\omega')^{2}+\omega_{D}^{2}}
\delta({\bf r}-{\bf r'}).
\end{eqnarray}
Here $\omega$ means the Matsubara frequency and $\psi_{n}$ denotes the 
scattered state. Subsequently, the strong-coupling gap equation can be easily 
obtained.$^{5}$
Note that the spatial part of the phonon Green's function 
$D({\bf r}-{\bf r'},\omega-\omega')$ becomes the Dirac delta function,
since the phonon frequency does not depend on the momentum.
Accordingly, the electron-phonon interaction coupling constant $\lambda$ is given by
\begin{equation}
\lambda=N_{o}<V_{nn'}(0,0)>=N_{o}{I_{o}^{2}\over M\omega_{D}^{2}}<\int
|\psi_{n}({\bf r})|^{2} |\psi_{n'}({\bf r})|^{2}d{\bf r}>.
\end{equation}
This result agrees with the BCS theory with a point interaction $V\delta({\bf r}_{1}-{\bf r}_{2})$, i.e.,
\begin{equation}
\lambda_{eff}=N_{o}V <\int|\psi_{n}({\bf r})|^{2} |\psi_{n'}({\bf r})|^{2}d{\bf r}>,
\end{equation}
where $V= I_{o}^{2}/ M\omega_{D}^{2}$.

Note that in the presence of impurities, the correlation function has a 
free-particle form for $t<\tau$ (scattering time) and a diffusive
form for $t>\tau$.$^{24}$ 
As a result, for $t>\tau$ (or $r>\ell$), 
one finds$^{25}$ 
\begin{eqnarray}
R &=& \int_{t>\tau} |\psi_{n}({\bf r})|^{2}|\psi_{n'}({\bf r})|^{2}d{\bf r} \nonumber\\ 
&=& \sum_{\vec q}|<\psi_{n}|e^{i{\vec q}\cdot {\bf r}}|\psi_{n'}>|^{2}_{AV}\nonumber\\
&=&\sum_{\pi/L<\vec q<\pi/\ell}{1\over 2\pi\hbar N_{o}D{\vec q}^{2}}\\
&=&{3\over 2(k_{F}\ell)^{2}}(1-{\ell\over L}).
\end{eqnarray}
Here $\ell$ is the mean free path and $L$ is the inelastic diffusion length.
Whereas the contribution from the free-particle-like density correlation
for $t<\tau$ is$^{5,25}$  
\begin{eqnarray}
V_{nn'} &=& V \int_{t<\tau} |\psi_{n}({\bf r})|^{2}|\psi_{n'}({\bf r})|^{2}d{\bf r} \nonumber\\ 
&\cong& V [1-{3\over (k_{F}\ell)^{2}}(1-{\ell\over L})].
\end{eqnarray}
Since the phonon-mediated interaction is retarded for 
$t_{ret}\sim 1/\omega_{D}$, only the free-particle-like density correlation
contributes to the pairing matrix element.
Thus, we obtain
\begin{eqnarray}
\lambda&=&N_{o}V [1-{3\over (k_{F}\ell)^{2}}(1-{\ell\over L})]\nonumber\\
&=&\lambda_{o} [1-{3\over (k_{F}\ell)^{2}}(1-{\ell\over L})].
\end{eqnarray}
Here $\lambda_{o}$ is the BCS $\lambda$ for the pure system.
Subsequently, one finds 
\begin{eqnarray}
\lambda_{tr} &=&2\int{\alpha_{tr}^{2}(\omega)F(\omega)\over \omega}d\omega\nonumber\\ 
&\cong&  \lambda_{o} [1-{3\over (k_{F}\ell)^{2}}(1-{\ell\over L})]\nonumber\\
&=& \lambda_{o} [1-{3\over (k_{F}\ell)^{2}}].
\end{eqnarray}
We have used the fact that $L$ is effectively infinite at $T=0$.
Note that the weak localization correction term is the same
as that of the conductivity.

\section{Explanation of the Mooij Rule }

The high temperature resistivity is then
\begin{eqnarray}
\rho_{ph}(T)&\cong&{2\pi mk_{B}T \over ne^{2}\hbar}\lambda\nonumber\\
&\cong& {2\pi mk_{B}T \over ne^{2}\hbar} \lambda_{o}[1-{3\over (k_{F}\ell)^{2}}].  
\end{eqnarray} 
On the other hand, the conductivity and the residual resistivity are given by
\begin{eqnarray}
\sigma&=&\sigma_{B}[1-{3\over (k_{F}\ell)^{2}}(1-{\ell\over L})],
\end{eqnarray}
and
\begin{equation}
\rho_{o}= {1\over  \sigma_{B}[1-{3\over (k_{F}\ell)^{2}}(1-{\ell\over L})]},
\end{equation}
where $\sigma_{B}=ne^{2}\tau/m$.
According to Matthiessen's rule, we may add both resistivities,
\begin{eqnarray}
\rho&\cong&\rho_{o}+\rho_{ph}(T)\nonumber\\
&=&
 {1\over  \sigma_{B}[1-{3\over (k_{F}\ell)^{2}}(1-{\ell\over L})]}
+{2\pi mk_{B}T \over ne^{2}\hbar} \lambda_{o}[1-{3\over (k_{F}\ell)^{2}}].  
\end{eqnarray} 

As the disorder parameter $1/k_{F}\ell$ is increasing, the system is more
disordered and the residual resistivity is getting higher. It is remarkable that the slope of the high temperature resistivity is decreasing concomitantly,
in good agreement with experiment. Note that the slope varies as
$\sim 1/(k_{F}\ell)^{2}$. This point has not been noticed before.
When $1/k_{F}\ell$ becomes comparable to $\sim 1$, the magnitude and slope of 
$\rho_{ph}(T)$ is becoming too small. In that case, only the residual 
resistivity will play an important role. 
Therefore, the  observed  negative TCR may be understood from the residual 
part. With decreasing $T$, 
since the inelastic diffusion length $L$ increases,
the residual resistivity will also increase, leading to the negative TCR.

Now we calculate Eq. (21) numerically to see the detailed temperature 
dependence of the resistivity of disordered systems.
Figure 3 shows the resistivity as a function of temperature.
We used $k_{F}=0.8\AA^{-1}$, $n=k_{F}^{3}/3\pi^{2}$,   
and $\lambda=0.5$.  
Since it is difficult to evaluate $k_{F}\ell$ up to a factor of 2,$^{26}$
we assume that $\rho=100\mu\Omega cm$ corresponds to $k_{F}\ell=3.2$.
We also used $L=\sqrt{D\tau_{i}}=\sqrt{\ell}\times 350/T (\AA)$.
Here $D$ is the diffusion constant and $\tau_{i}$ denotes the inelastic 
scattering time.
For low temperatures $\tau_{i}$ is determined by electron-electron scattering
while for high temperatures it is determined by the electron-phonon scattering. 
Since we are interested in rather high temperatures, 
we assumed $\tau_{i}\sim T^{-1}$ corresponding to the electron-phonon 
scattering.  
Considering the crudeness of our calculation, the overall behavior is in good 
agreement with experiment.

\section{\bf Discussion} 

It is clear that weak localization effect on the electron-phonon interaction
needs more theoretical and experimental studies. In particular, weak 
localization effect on the attenuation of a sound wave, shear modulus, 
thermal resistance, and a shift in phonon frequencies will be very interesting.
Since superconductivity is also caused by the electron-phonon interaction,
comparative study of  the normal and superconducting properties of the metallic
samples will be beneficial. There is already compelling evidence that this
is the case. For instance, Testardi and his coworkers$^{27-30}$ found the 
universal correlation of $T_{c}$ and the resistance ratio. They also found that
decreasing $T_{c}$ is accompanied by the decrease of the thermal electrical
resistivity.$^{27}$ 

Note that this study may provide a means of probing the phonon-mechanism
in exotic superconductors, such as, heavy fermion superconductors,
organic superconductors, fullerene superconductors,  and high $T_{c}$ cuprates. 
For superconductors caused by the electron-phonon interaction we expect the
following behavior. As the electrons are weakly localized by impurities or radiation damage, 
the electron-phonon interaction is weakened. As a result, both $T_{C}$
and TCR are decreasing at the same rate. When $\lambda$ is approaching zero,
both $T_{c}$ and TCR drops to zero almost simultaneously. 
When this happens we may say that 
the electron-phonon interaction is the origin of the pairing in the 
superconductors.
This behavior was already confirmed in A15 superconductors$^{27-30}$ and Ternary superconductors.$^{31}$
More details will be published elsewhere.

\section{\bf Conclusion} 

It is shown that weak localization decreases 
both the conductivity and the electron-phonon interaction at the same rate 
and thereby leads to  the Mooij rule. 
As the residual resistivity is increasing due to weak localization, so
the thermal electrical resistivity is decreasing,
producing the decrease of TCR. When the electron-phonon interaction is
near zero, only the residual resistivity is left and therefore
the negative TCR is obtained. Matthiessen's rule seems to be intact
to a large extent even in highly disordered systems.
This study may provide a means of probing the phonon-mechanism in exotic
superconductors, such as, heavy fermion superconductors, organic 
superconductors, fullerene superconductors, and high $T_{c}$ superconductors.

\vspace{1pc}

\centerline{\bf ACKNOWLEDGMENTS}

YJK is grateful to Prof. Bilal Tanatar for discussions and
encouragement. M. Park thanks the FOPI at the University of Puerto Rico-Humacao
for release time.

\vfill\eject

{\bf Table I.} Comparison of $\lambda_{tr}$ and $\lambda$ as given in Ref. 20.

\vspace{2pc}
\hspace{2pc}

\begin{tabular}{lrrlrr} \hline \hline
{Metal}\hspace{2pc}  & $\lambda_{tr}$\hspace{2pc} & \hspace{2pc} {$\lambda$}& 
\hspace{4pc}{Metal}\hspace{2pc}  & $\lambda_{tr}$\hspace{2pc} & \hspace{2pc} {$\lambda$}  \\ \hline

Li \hspace{2pc} & .40 \hspace{2pc} & \hspace{1pc} .41$\pm$.15 & \hspace{4pc} Na  \hspace{2pc} & .16 \hspace{2pc} & \hspace{1pc} .16$\pm$.04\\ 

K \hspace{2pc}  & .14 \hspace{2pc} & \hspace{1pc} .13$\pm$.03 & \hspace{4pc} Rb \hspace{2pc} & .19 \hspace{2pc} & \hspace{1pc} .16$\pm$.04\\ 

Cs \hspace{2pc} & .26 \hspace{2pc} & \hspace{1pc} .16$\pm$.06 & \hspace{4pc} Mg \hspace{2pc} & .32 \hspace{2pc} & \hspace{1pc} .35$\pm$.04\\ 

Zn \hspace{2pc} & .67 \hspace{2pc} & \hspace{1pc} .42$\pm$.05 & \hspace{4pc} Cd \hspace{2pc} & .51 \hspace{2pc} & \hspace{1pc} .40$\pm$.05\\ 

Al \hspace{2pc} & .41 \hspace{2pc} & \hspace{1pc} .43$\pm$.05 & \hspace{4pc} Pb \hspace{2pc} & 1.79 \hspace{2pc} & \hspace{1pc} 1.55\\ 

In \hspace{2pc} & .85 \hspace{2pc} & \hspace{1pc} .805 & \hspace{4pc} Hg \hspace{2pc} & 2.3 \hspace{2pc} & \hspace{1pc} 1.6\\ 

Cu \hspace{2pc} & .13 \hspace{2pc} & \hspace{1pc} .14$\pm$.03 & \hspace{4pc} Ag \hspace{2pc} & .13 \hspace{2pc} & \hspace{1pc} .10$\pm$.04\\ 

Au \hspace{2pc} & .08 \hspace{2pc} & \hspace{1pc} .14$\pm$.05 & \hspace{4pc} Nb \hspace{2pc} & 1.11 \hspace{2pc} & \hspace{1pc} .9$\pm$.2\\ \hline \hline

\end{tabular}

\vfill\eject

\begin{figure}
\caption{ The temperature coefficient of resistance $\alpha$ versus resistivity for bulk alloys (+), thin films ($\bullet$), and amorphous (X) alloys. Data are from Mooij, Ref. 7. }
\end{figure}

\begin{figure}
\caption{  Resistivity versus temperature for Ti and TiAl alloys containing 
0, 3, 6, 11, and 33\% Al. Data are from Mooij, Ref. 7. }
\end{figure}
\begin{figure}
\caption{  Calculated resistivity versus temperature for $k_{F}\ell=$ 15, 5, 
3.4, 2.8, 2.4, and 2.3. \hspace{2pc} }
\end{figure}

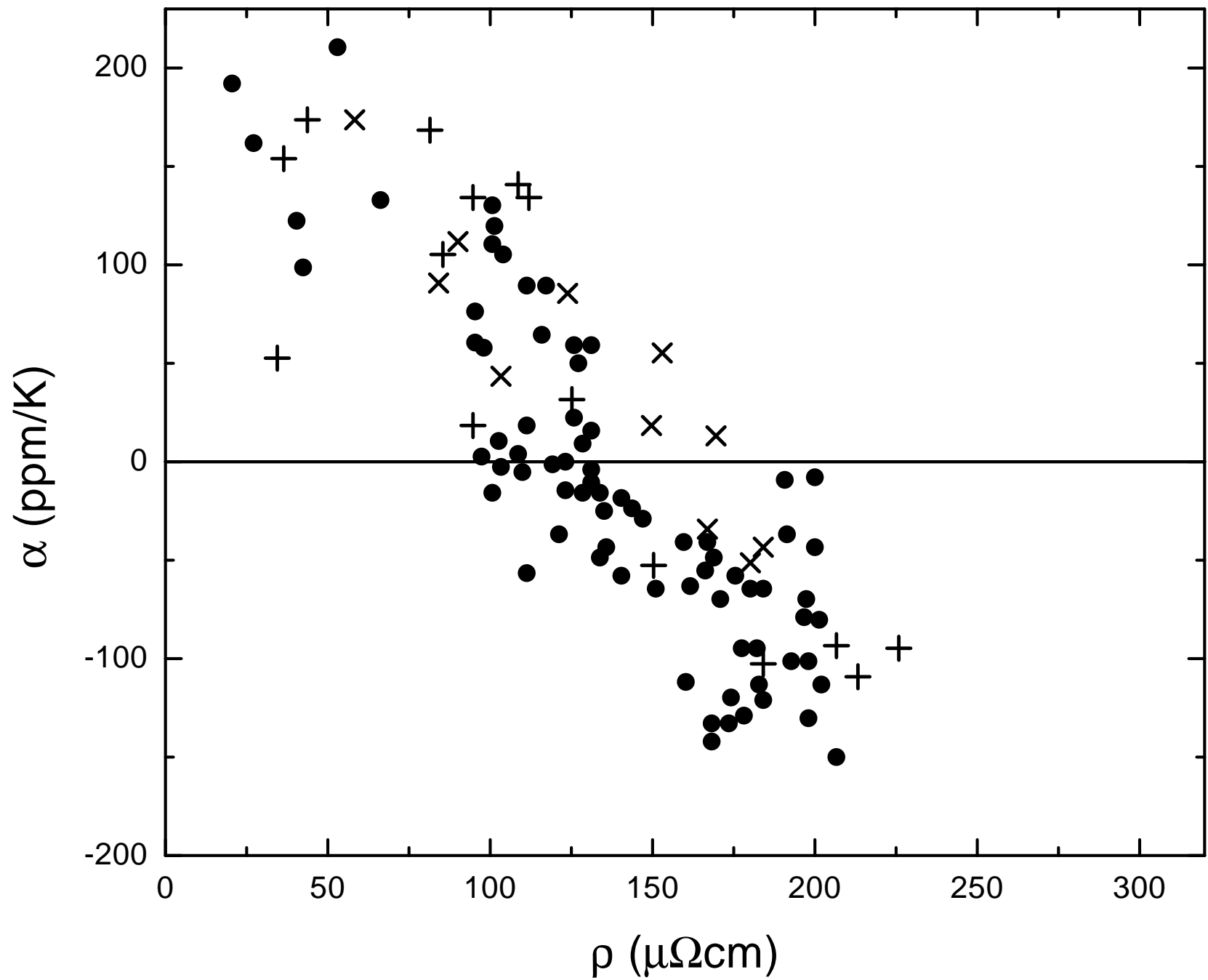
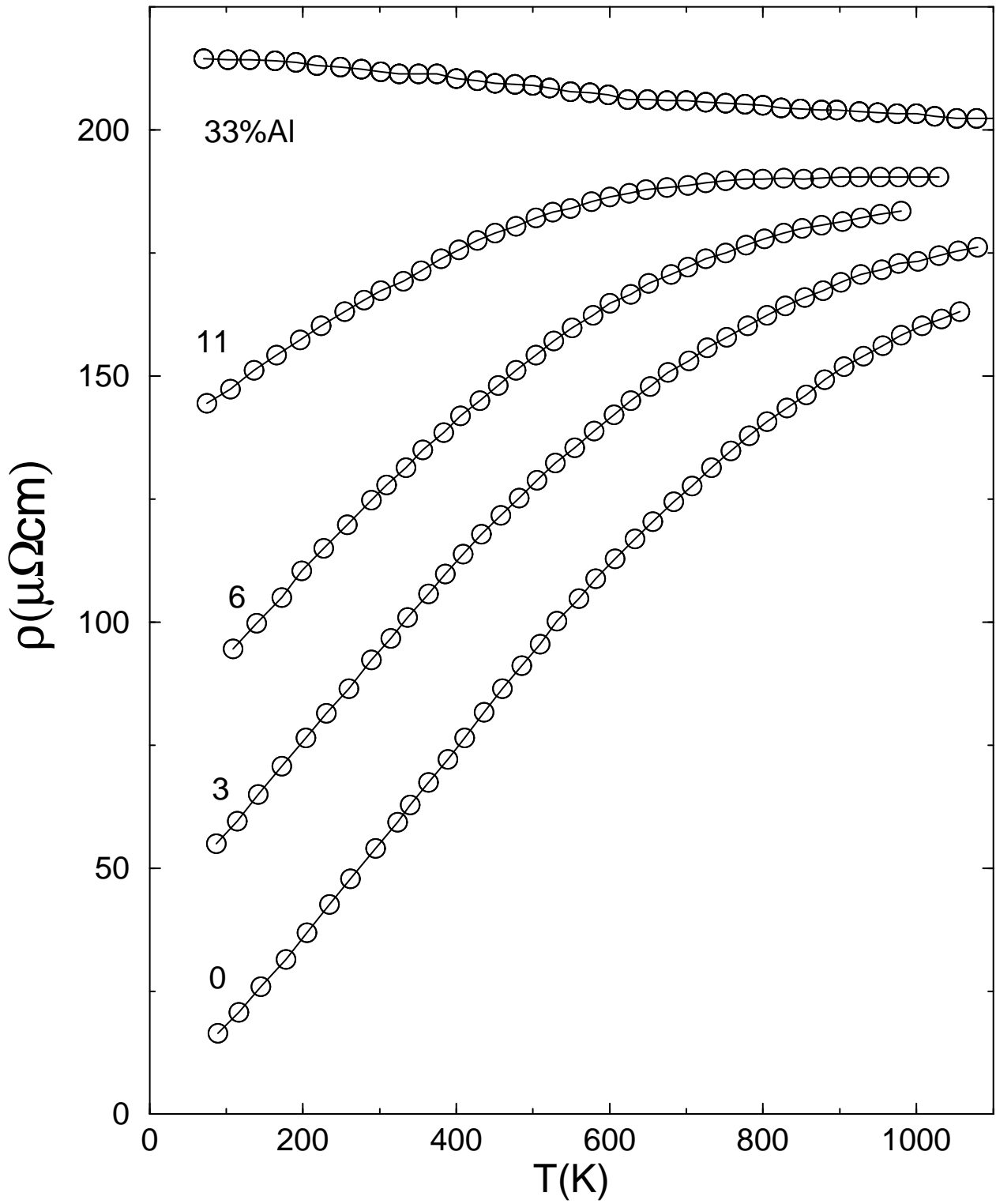
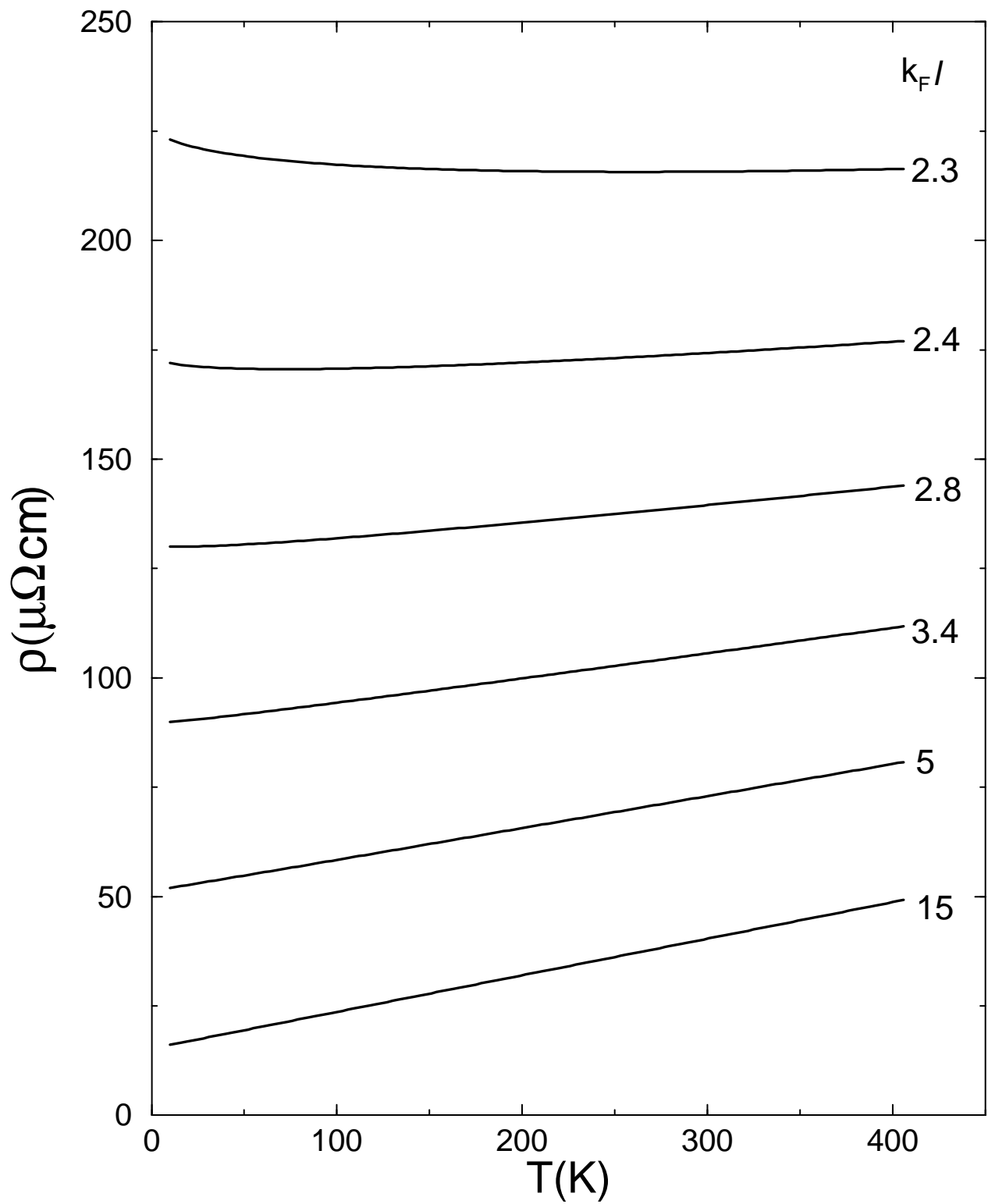
\end{document}